\documentstyle[12pt,epsfig]{elsart}

\date{}
\def\Journal#1#2#3#4{{#1} {\bf #2}, #3 (#4)}
\def\NCI{\em Nuovo Cimento}
\def\NIM{\em Nucl. Instrum. Methods}
\def\NIMA{{\em Nucl. Instrum. Methods} A}

\def\NPA{{\em Nucl. Phys.} A} 
\def\APB{{\em Acta Pol.} B}
 
\def\PLB{{\em Phys. Lett.} B}
\def\PRP{{\em Phys. Rep.}}
 
\def\JPG{\em Journal of Physics G}
 
\def\PR{\em Phys. Rev.} 
 
\def\PRC{{\em Phys. Rev.} C} 
\def\ZPC{{\em Z. Phys.} C} 
 
\def\PNPP{\em Prog. Nucl. Part. Phys.} 
\def\PTP{\em Prog. Theo. Phys.}

\def\PAN{{\em Physics of Atomic Nuclei}}

\def\ra{\rightarrow} 
\def\be{\begin{equation}} 
\def\ee{\end{equation}}
\def\bea{\begin{eqnarray}} 
\def\eea{\end{eqnarray}} 

\newcommand{\obb}{0\mbox{$\nu\beta\beta$ - decay} } 
\newcommand{\zbb}{2\mbox{$\nu\beta\beta$ - decay} }

\newcommand{\bpbp}{\mbox{$\beta^+\beta^+$} }
\newcommand{\ecec}{\mbox{$EC/EC$} }

\newcommand{\bec}{\mbox{$\beta^+/EC$} }
\newcommand{\ton}{\mbox{$T_{1/2}$}}

\newcommand{\delm}{\mbox{$\Delta m^2$} }

\newcommand{\nel}{\mbox{$\nu_e$} }

\newcommand{\neu}{neutrino }

\newcommand{\ema}{\mbox{$\langle m_{\nu_e} \rangle $}}

\newcommand{\moeh}{\mbox{$^{100}$Mo }}

\newcommand{\tinhz}{\mbox{$^{112}$Sn }}
\newcommand{\cdhz}{\mbox{$^{112}$Cd }}
\newcommand{\tinun}{\mbox{$^{122}$Sn }}
\newcommand{\tinhv}{\mbox{$^{124}$Sn }}
\newcommand{\tehv}{\mbox{$^{124}$Te }}
 
\newcommand{\ndhf}{\mbox{$^{150}$Nd} }

\newcommand{\bnel}{\mbox{$\bar{\nu}_e$} }

\begin{document}
\begin{frontmatter} 
\title{A search for various double beta decay modes of tin isotopes} 
\author{J. Dawson, R. Ramaswamy, C. Reeve, J. R. Wilson, K. Zuber} 
\address{Dept. of Physics and Astronomy, University of Sussex\\  
Falmer, Brighton BN1 9QH, UK} 
\begin{abstract} 
For the first time an extensive search for various double beta decay modes of \tinhv and \tinhz has been performed. A total exposure of 43.29 kg$\times$days has been accumulated. New half-life limits of \tinhv into excited states of $^{124}$Te have been obtained; the lower half-life limit for the first excited 2$^+$ state  at 602.7\,keV is \ton $> 3.1 \times 10^{18}$yrs (90 \% CL) and for the first excited 0$^+$ state \ton  $> 7.7 \times 10^{18} $yrs (90 \% CL). For the very first time, ground state and excited state transitions of \tinhz have been experimentally explored. The obtained half-life limits for \ecec and \bec into the first excited 2$^+$ state of $^{112}$Cd are both \ton$>1.4\times 10^{18}$yrs (90 \% CL). A resonance enhancement in the decay rate for $0\nu$ \ecec  might be expected for the 0$^+$-state at 1870.9\,keV due to degeneracy with the \tinhz ground state. No signal was found resulting in a lower half-life limit of \ton  $> 1.6 \times 10^{18}$ yrs (90 \% CL) for this decay. As all the excited state searches are based on gamma-lines, all half-life limits apply for both neutrino and neutrino-less modes. Neutrinoless ground state transitions were searched for in the \ecec and \bec mode and a limit of \ton  $> 1.5 \times 10^{18}$ yrs (90 \% CL) was obtained for \ecec decays of $^{112}$Sn, whilst the \bec mode results are inconclusive.
\end{abstract} 
\end{frontmatter} 

\section{Introduction} 
Over the last few years striking evidence has arisen for a non-vanishing neutrino rest mass (for reviews see Refs.~\cite{zub98,samoil}). This evidence comes from neutrino oscillations experiments which are not able to measure absolute neutrino masses, because their results depend only on the differences of masses-squared, \delm = $m_i^2- m_j^2$, with $m_i,m_j$ as the masses of two neutrino mass eigenstates. Various scenarios have been proposed to describe these mass eigenstates on the absolute neutrino mass scale. If neutrino masses are in the region larger than 0.1\,eV they will be almost degenerate, otherwise hierarchical structures can be observed (normal or inverted). A very sensitive process to probe the absolute neutrino mass is 0\mbox{$\nu\beta\beta$ - decay}. Double beta decay was first discussed by M. Goeppert-Mayer~\cite{goe35} in the form of 
\be
(Z,A) \ra (Z+2,A) + 2 e^- + 2 \bar{\nu_e} \quad (\zbb).
\ee
This process can be seen as two simultaneous neutron decays. Shortly after the classical papers of Majorana~\cite{maj37} discussing a 2 - component neutrino, Racah~\cite{rac37} and Furry discussed another decay mode in the form of \cite{fur39}
\be
\label{proc0nu}
(Z,A) \ra (Z+2,A) + 2 e^-  \quad (\obb).
\ee
In contrast to neutrino oscillations which violate individual flavour lepton number, but keep total lepton number conserved, \obb\ violates total lepton number by two units. This process is forbidden in the Standard Model. It can be seen as two subsequent steps (``Racah - sequence'') 
\bea
(Z,A) \ra &(Z+1,A)& + e^- +  \bar{\nu_e} \nonumber\\
&(Z+1,A)& + \nel \ra (Z+2,A) + e^- 
\eea
The quantity measured in \obb is called effective Majorana \neu mass and given for light neutrinos by 
\be 
\label{eq:ema}\ema = \mid \sum_i U_{ei}^2 m_i \mid =  \mid \sum_i \mid U_{ei} \mid^2 e^{2i\alpha_i} m_i \mid 
\ee 
which can be written in case of CP-invariance $e^{2i\alpha_i} = 0, \pi$ as 
\be
\ema = \mid  m_1 \mid U_{e1}^2 \mid \pm m_2 \mid U_{e2}^2 \mid \pm m_3 \mid U_{e3}^2 \mid \mid
\ee 
The experimental signal of \obb\ is two electrons in the final state, whose energies add up to the Q-value of the nuclear transition, while for the \zbb the sum energy spectrum of both electrons will be continuous. The total decay rates, and hence the inverse half-lives, are a strong function of the available Q-value. The rate of \obb scales with $Q^5$ compared to a $Q^{11}$-dependence for 2\mbox{$\nu\beta\beta$ - decay}. Therefore isotopes with a high Q-value (above about 2\,MeV) are normally considered for experiments.  This restricts the candidates to eleven promising isotopes. For recent reviews on double beta decay see Refs.~\cite{ell04,zub06,avi07}.\par
As an alternative to the ground-state transitions, decays into excited 0$^+$ and 2$^+$-states of the daughter can be used for the search as well. This might shed some light on new physics and provide some new information on nuclear matrix element calculations. Observations of \zbb transitions into the first excited 0$^+$-state for \moeh and \ndhf have been reported \cite{bar95,bar04,arn07,bar07}. Despite a reduced Q-value for such transitions a search using the de-excitation gammas alone, or in coincidence with the two electrons, provides a clear signal of the decay.\par
Recently there has been a revived interest to study double electron capture for neutrino mass searches. Three different decay channels can be considered combining electron capture (EC) and positron emission:
\bea
(Z,A) &\ra& (Z-2,A) + 2 e^+ + (2 \nel) \quad (\bpbp)\\
e^- + (Z,A) &\ra& (Z-2,A) + e^+ + (2 \nel) \quad (\bec)\\ 
2 e^- + (Z,A) &\ra& (Z-2,A) + (2 \nel)  \quad\quad\quad (\ecec)
\eea
In particular, the \bec mode shows an enhanced sensitivity to right handed weak currents \cite{hir94}. The experimental signatures of the decay modes involving positrons in the final state are promising because of two or four 511\,keV photons. Despite this nice signature, they are less often discussed in literature, because for each generated positron the available Q-value is reduced by $2m_ec^2$, which leads to much smaller decay rates than in comparable \obb due to reduced phase space. Hence, for \bpbp{}-decay to occur, the Q-values must be at least 2048\,keV. Only six isotopes are know to have such a high Q-value. The full Q-value is only available in the \ecec mode, which is harder to detect. In a neutrinoless \ecec to the ground state of the daughter a monoenergetic internal bremsstrahlung photon has to be emitted. This requires a K- and an L-shell capture. Recently a resonant enhancement in the decay rate of radiative \ecec has been explored; if the initial and final states are degenerate \cite{suk04}, the de-excitation photons will serve as a signal. Thus a wide variety of possible gamma emissions exist.\par
\begin{figure}
\begin{center}
\epsfig{file=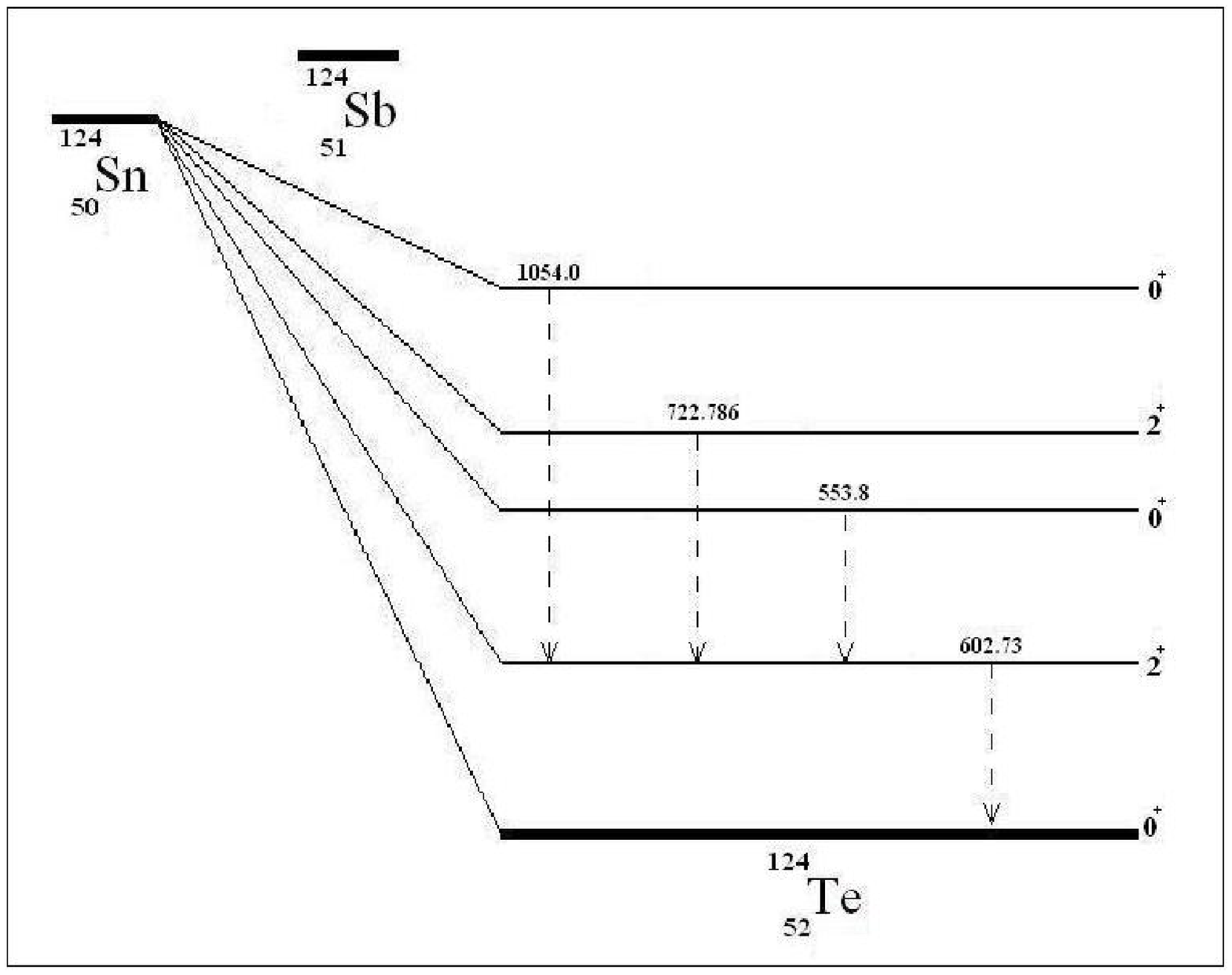,width=6cm}\hspace{1cm}
\epsfig{file=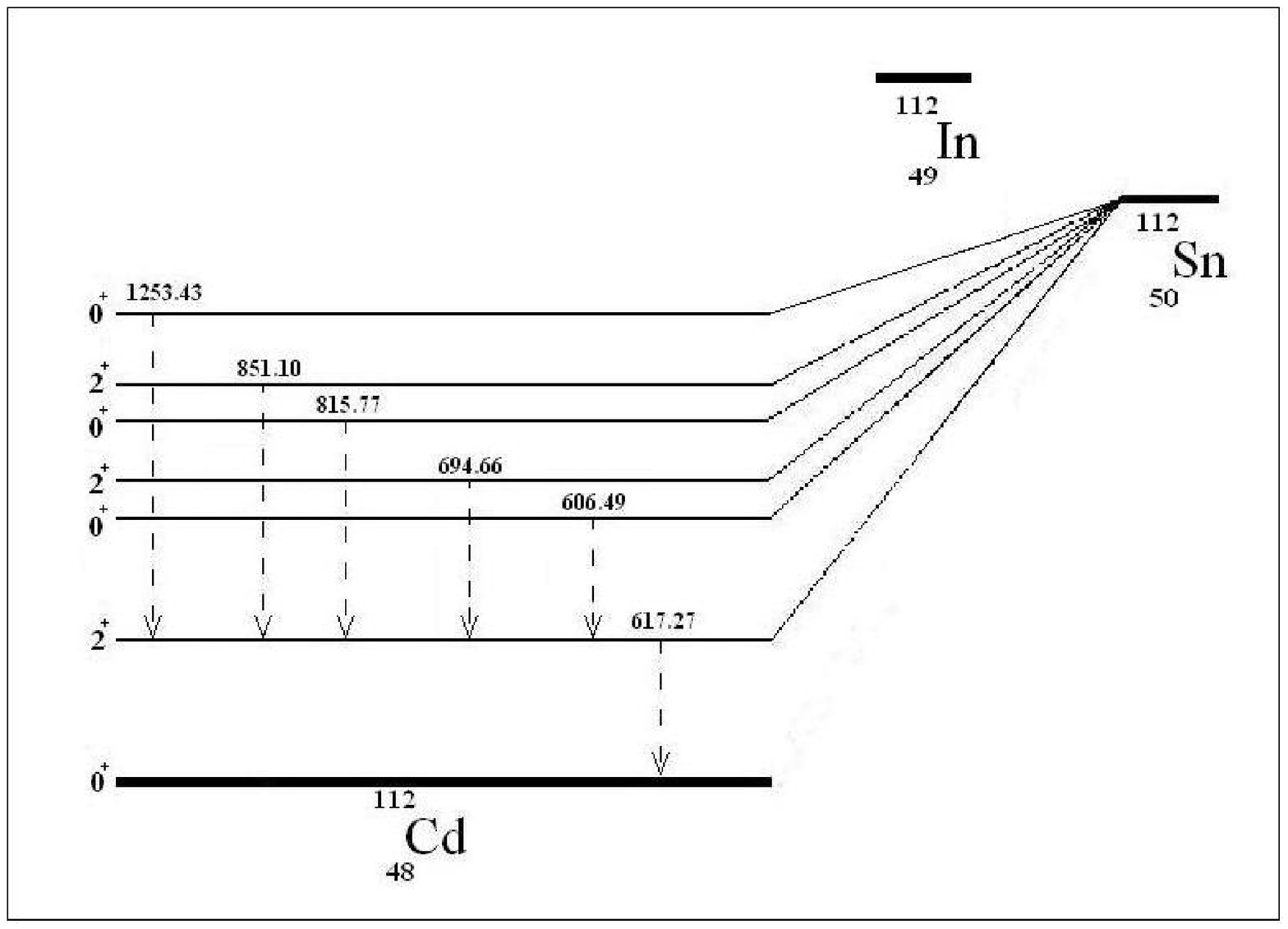,width=6cm}
\end{center}
\caption{\label{pic:levels} Double beta decay schemes of \tinhv and $^{112}$Sn. Left: The \tinhv scheme with the corresponding 0$^+$- and 2$^+$-states. The dominant de-excitation chains of all higher levels proceed via the 2$^+_1$-state and result in the emission of a 603\,keV gamma. Right:
Decay schema of $^{112}$Sn.}
\end{figure}

This paper explores double beta transitions of tin isotopes. Tin contains three double beta emitters, \tinun and \tinhv in the two electron mode and \tinhz for \bec and \ecec decays. The Q-values of the transitions for the  three isotopes are 366\,keV, 2287\,keV and 1919\,keV and the natural abundances are 4.63 \%, 5.79 \% and 0.97 \% respectively. As there is no excited state of interest for \tinun decay we focus on the decays
\bea 
\tinhv &\ra& \tehv + 2e^- + (2 \bnel) + \gamma \\
\ecec + \tinhz &\ra& \cdhz + (2 \nel) + \gamma \\
\bec + \tinhz &\ra& \cdhz + (2 \nel) + \gamma 
\eea
The decay schemes of \tinhv and \tinhz are shown in Fig.~\ref{pic:levels}. Astonishingly, despite the fact that \tinhv is one of the eleven isotopes with a Q-value larger than 2\,MeV there is currently no experimental suggestion for a large scale experiment. No theoretical predictions for half-lives exist for \tinhz and $^{122}$Sn, calculations are reported for ground state and excited states transitions in \tinhv \cite{aun96,cau99}. Ground state limits for \tinhv and \tinun in the range of $5-6\times10^{13}$\,yr have been deduced~\cite{fre52}, while for \tinhv there is a lower limit of $2.4 \times 10^{17}$\,yrs for the neutrinoless-mode \cite{kal52}.	A search for the decay into first excited 0$^+$ and 2$^+$-states of \tehv is reported in \cite{nor87}.\par
The presence of 511\,keV gammas in the \bec mode allows a search for a ground state transition in $^{112}$Sn.  The neutrinoless ground state transition of \ecec can be explored by searching for the monoenergetic internal brems-strahlung gamma at $Q-E_K-E_L$ = 1888.5\,keV, with $E_K, E_L$ the K- and L-shell binding energy of $^{112}$Cd. Also interesting is the fact that there might be a degeneracy between \tinhz and an excited 0$^+$ state at 1870.9\,keV in \mbox{$^{112}$Cd}, hence fulfilling the resonance enhancement condition for neutrinoless \ecec \cite{bar07}. Gamma line energies were taken from \cite{toi} and the latest atomic mass determinations from \cite{wap03}.
\section{Experimental Setup}
\begin{figure}
\begin{center}
\epsfig{file=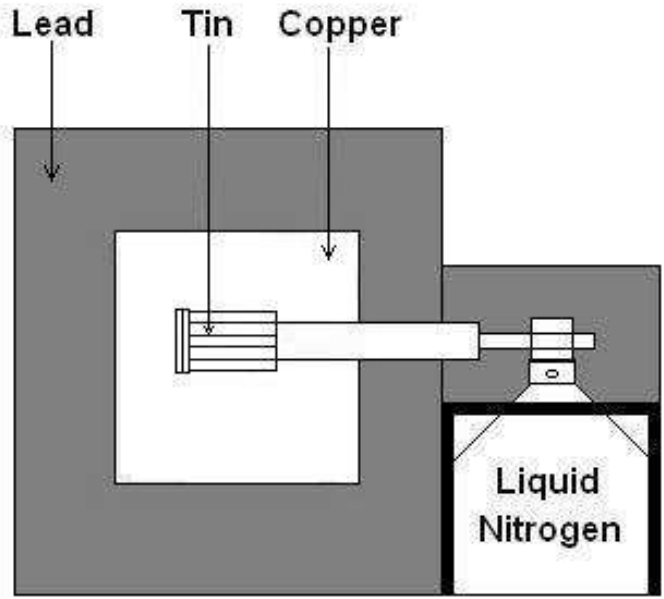,width=10cm}
\end{center}
\caption{\label{pic:setup} Schematic drawing of the experimental set-up used. The tin source is mounted on the encapsulation of a Ge-detector to optimise the efficiency for gamma detection.}
\end{figure}

\begin{figure}
\begin{center}
\epsfig{file=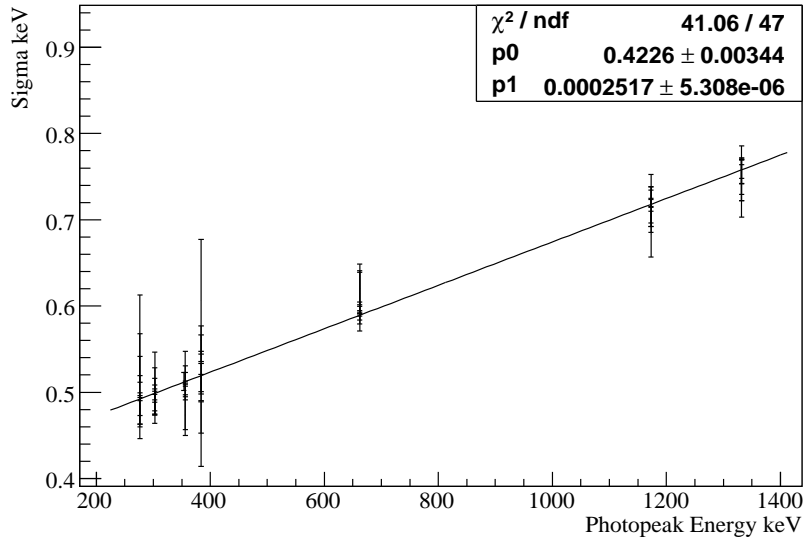,width=12cm,height=8cm}
\end{center}
\caption{\label{pic:reso} Energy resolution (sigma) as function of energy obtained with standard calibration sources as described in the text.}
\end{figure}

The setup is schematically shown in Fig.~\ref{pic:setup}. For the search, a 72~cm$^3$ Ge-detector was used. The tin with guaranteed 99.95\,\% purity was mounted in two different ways around the detector. Two concentric rings consisting of 18 tin pieces, each with dimensions 7.25 $\times$ 1.4 $\times$ 0.35 cm$^3$, were mounted around the cylindrical surface of the Ge-detector. The tin pieces were held in place by a thin band of copper braid and further supported by the surrounding copper shielding. The front part of the Ge-detector was covered by two sheets of tin, their size determined by the size of the Ge-detector and the tin ingot. The total amount of tin used was 1.24\,kg. The whole set-up was surrounded by 10~cm of clean copper and 10--20\,cm of lead, slightly dependent on the direction. The experiment was located in a surface laboratory without neutron shielding so some activity from cosmic sources is expected. Due to the presence of background lines from the Uranium and Thorium decay chains, in-situ calibrations could be performed. Independent calibrations were performed with $^{133}$Ba, $^{137}$Cs, $^{22}$Na and $^{60}$Co. Measurement runs were taken in units of hours to optimise data taking. Spectra were recorded with an ORTEC MCA card of 11 bit resolution. Efficiencies of the setup for the gamma rays of interest were studied with the help of GEANT4 Monte Carlo simulations. To check the simulation, detailed calibration runs with $^{57}$Co, $^{137}$Cs and $^{133}$Ba source were performed for a number of different source positions. The region of interest for the search contains gamma energies in the range of 500--1900\,keV. The energy resolution as a function of energy is shown in Fig.~\ref{pic:reso}. 
\begin{table}[htbp]
\begin{center}
{\small
\begin{tabular}{|c|c|c|c|}
\hline
Run  & Measuring time & Energy range & Configuration \\
& (hrs) & (keV) &  \\ \hline
2 & 89.44 & 5--1550  & 1 cyl. ring, 14.5\,cm long, 1 front layer\\
3 & 167.00 & 5--1550 & 1 cyl. ring, 14.5\,cm long, 2 front layers\\
4 & 73.65 & 5--2020 & 2 cyl. rings, 7.25\,cm long, 2 front layers\\
5 & 95.38 & 5--2020 & 2 cyl. rings, 7.25\,cm long, 2 front layers\\
6 & 406.00 & 5--2020 & 2 cyl. rings, 7.25\,cm long, 2 front layers\\
\hline
\end{tabular}
}
\caption{\label{tab:runs} Experimental runs taken. Efficiencies were calculated for each different tin configuration separately. (Run 1 was used for commissioning purposes only).}
\end{center}
\end{table}
\section{Analysis}
Individual measurements were stored every hour to minimise data taking inefficiencies. The total measuring period was divided into subsets, called runs, their length determined by calibration measurements or optimisation of the data taking and shielding. They are listed in Tab.~\ref{tab:runs}.\par
\begin{figure}
\begin{center}
\epsfig{file=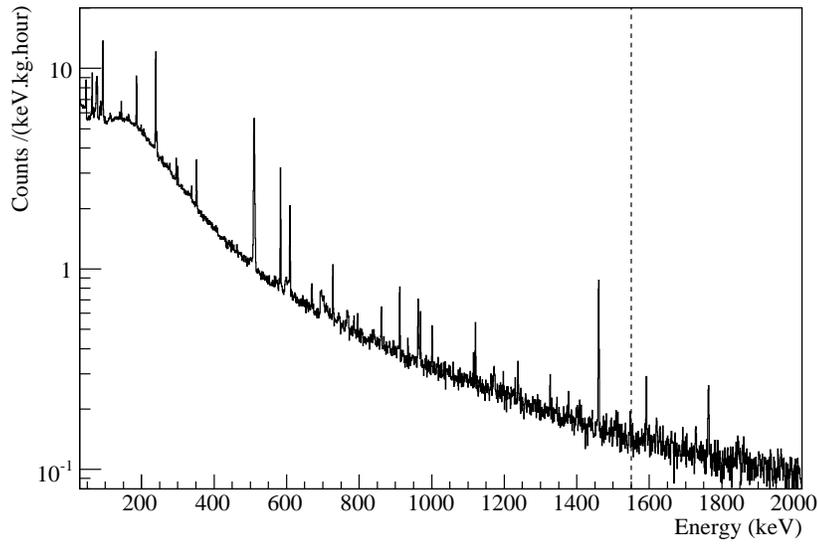,width=12cm,height=8cm}
\end{center}
\caption{\label{pic:totalspec} Total spectrum obtained. To the right of the dashed line, only runs 4--6 contribute to the spectrum. }
\end{figure}

The total measuring time of the obtained data is 831.47 hours corresponding to a total statistics of 43.29\,kg$\times$days, taking into account the slightly different masses in Run 2 and Run 3. However, note that only the last three runs contribute to the spectrum above 1550\,keV. The total spectrum obtained is shown in Fig.~\ref{pic:totalspec}. \par
\begin{figure}
\begin{center}
\epsfig{file=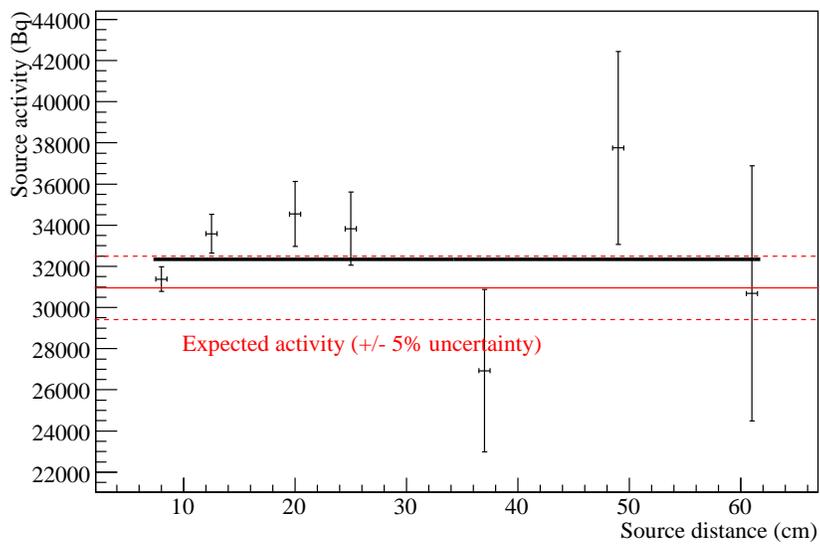,width=12cm,height=8cm}
\end{center}
\caption{\label{pic:eff} The activity derived for the $^{133}$Ba calibration source for a number of source positions (x-axis gives distance of source from front of detector). The red band shows the true source strength, known to $\pm$5\% accuracy, and the black line gives the average calculated value. }
\end{figure}

Detection efficiencies have to be determined for a variety of gamma rays involved in the decay modes. Thus, detailed GEANT4-based simulations have been performed modelling all parts of the detector, source and shielding. The accuracy of the simulation was verified through comparison of simulated events and data obtained with the three mentioned collimated calibration sources (The calculated activities for $^{133}$Ba are compared with the known value in Fig.~\ref{pic:eff}). The accuracy is better than 5 \% and limited by the uncertainties in the source strengths.\par
To determine the number of possible events for each predicted gamma emission, a Bayesian approach was used~\cite{pdg-stats}. A binned maximum likelihood fit was performed to determine the most likely values for the magnitude of a linear background component and the amplitude of a gaussian peak at each possible gamma energy. As well as the mean, the width of this gaussian was fixed using the calibrated resolution function and each fit was performed over a range of $\pm 30 \sigma$ around the known gamma energy. For some of the signals analysed, gamma lines from radioactive contaminants lie within the selected fit region. In these cases, an additional gaussian function with fixed mean and width was added to the fit, and the amplitude varied as a fit parameter. \par
\begin{figure}
\begin{center}
\epsfig{file=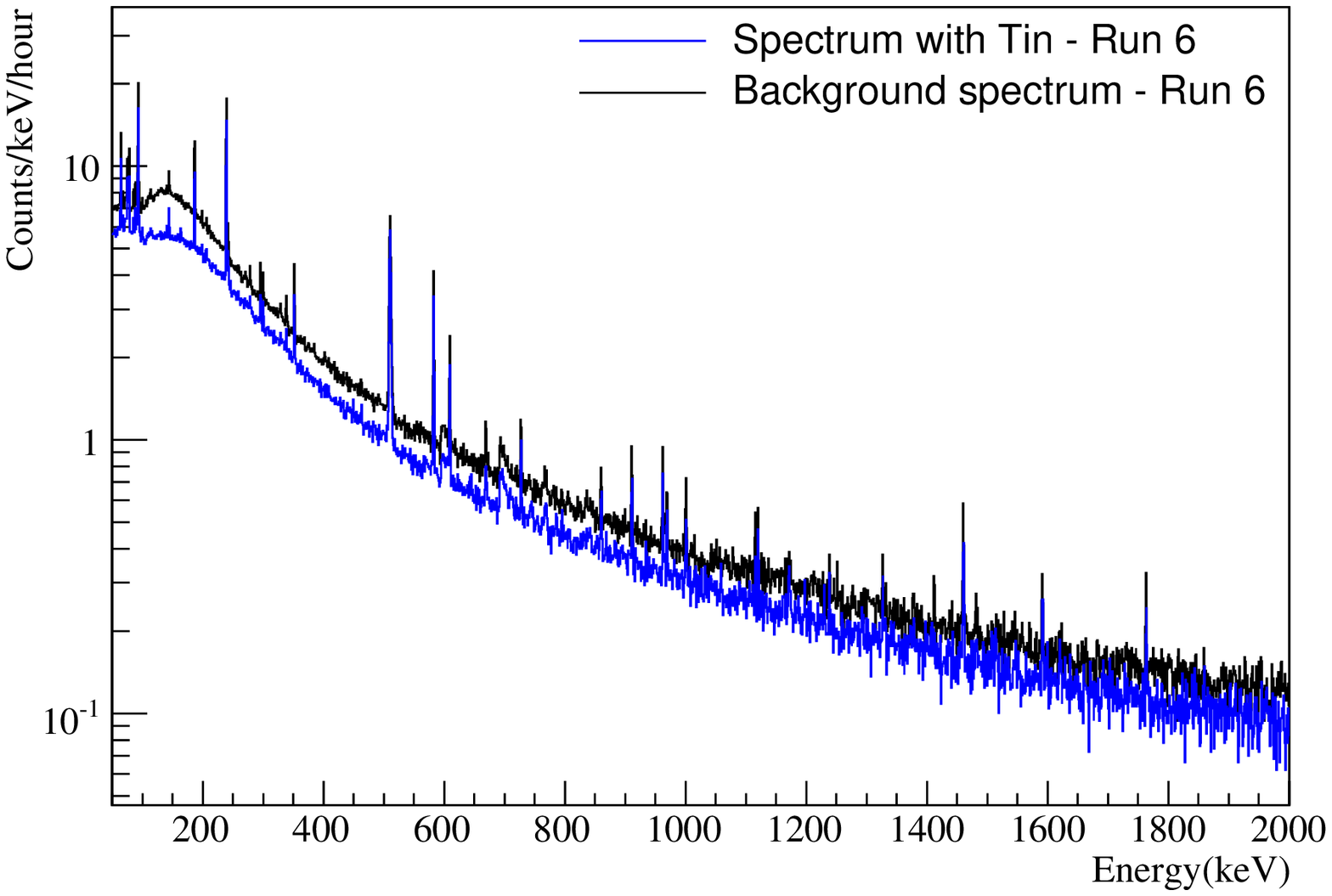,width=12cm,height=8cm}
\epsfig{file=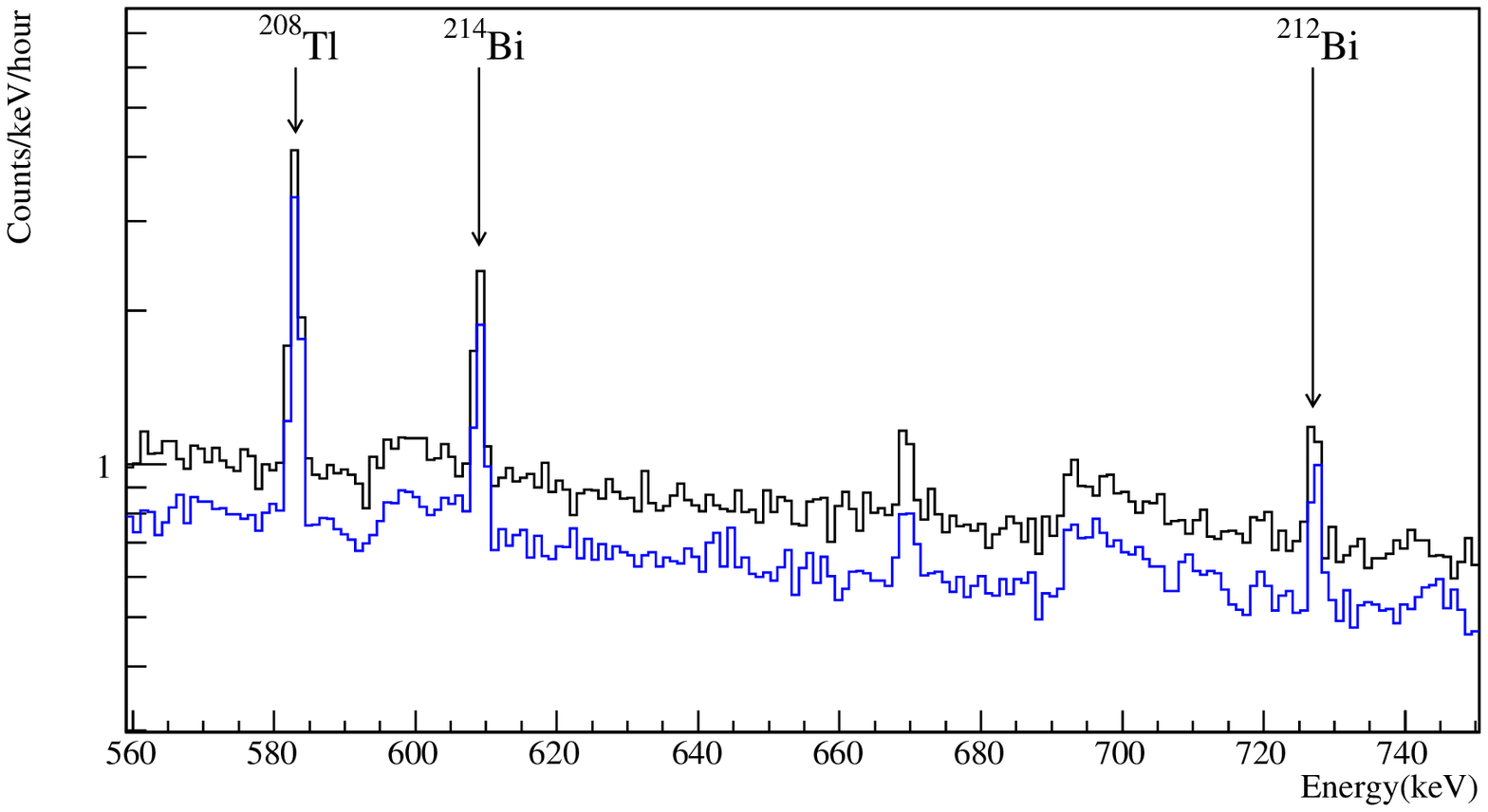,width=12cm,height=8cm}
\end{center}
\caption{\label{f:background} Run 6 data spectrum obtained with Tin compared to the spectrum obtained without tin present (ie. background only). The bottom plot shows the range 560--740\,keV in more detail, including the two features at 596 and 691\,keV due to fast neutron interactions on Ge. The peak at 669\,keV is thought to be from $^{63}$Zn or $^{60}$Zn, which are also produced by neutron irradiation of germanium\cite{entwistle}.}
\end{figure}

Some features of the background spectrum do not exhibit gaussian shapes; in particular the regions 590--610\,keV and 690--720\,keV shown in Fig.~\ref{f:background}. These features result from neutron interactions inside the detector producing de-excitation gamma-rays at 596\,keV and 691\,keV with additional contributions from nuclear recoil\cite{bunting,gete}. It should be noted that similar features can also be seen in the spectrum presented in Ref.~\cite{nor87}, though are not explicitly treated. Fig.~\ref{f:background} shows a spectrum obtained with the shielding setup specified for Run 6, but with the tin pieces removed, consisting of 537\,hours of data, compared to the data obtained with tin for Run 6. The tin spectrum contains less counts at low energies due to the shielding effect of the tin pieces on external backgrounds. To account for this difference, the total attenuation (including coherent scattering) of tin as a function of energy, $A(E)$, obtained from Ref.~\cite{xcom} was used to modify the background spectrum such that the counts per bin, $N_b$ became $N_b'$ via
\begin{equation} 
{N_b' = N_b\times\left(\left(1 - f\right) + f\times \exp\left(-A(E)\rho x_{ave}\right)\right)}
\label{e:atten}
\end{equation}
where $f$ is the fraction of background that penetrates the tin,\footnote{The tin does not cover the full solid angle of the detector and some of the background may arise inside the tin layer.} $x_{ave}$ is the average thickness of tin penetrated, and $\rho =  7.31$\,g\,cm$^{-3}$ is the density of tin. \par
For the four gamma lines coincident with these spectral features (602.7 and 695.0\,keV for $^{124}$Sn and 606.5 and 694.7\,keV for $^{112}$Sn), rather than fitting a linear background, the attenuated background spectrum was fitted to the data instead, with the parameters $f$ and $x_{ave}$ as variables in the fit. The amplitude of the gaussian peak was also allowed to float in the fit, with the mean and width fixed as before. The four fits resulted in consistent values of $f\approx 0.4$ and $x_{ave}\approx 1.3$\,cm and the fitted event numbers and half-life limits given in Tabs.~\ref{tab:exstate-tinhv} and \ref{tab:exstate-tinhz} (indicated by a \dag). Note that because the background was only measured with high statistics for the shielding configuration of Run 6, only the Run 6 data was analysed for these decay modes. \par
For both approaches, the fitted signal amplitude (ie. total number of signal events), and  uncertainty determined from the likelihood fit, $\theta_S \pm \delta_S$, was then used to derive a lower limit on the half-life for each decay mode using Eqn~\ref{e:thalflim}.
\begin{eqnarray}
T_{\rm half} &\le& \frac{\ln2 \sum_{i=1}^{i=N}{N_{\rm iso}^i t_{\rm live}^i \epsilon^i \zeta}}
{\sum_{i=1}^{i=N}\theta_s^i +\left(\sum_{i=1}^{i=N}\left(1.28\delta_s^i\right)^2\right)^{1/2}}\hspace{1cm}(\sum\theta_s^i > 0)\nonumber \\ 
T_{\rm half} &\le& \frac{\ln2 \sum_{i=1}^{i=N}{N_{\rm iso}^i t_{\rm live}^i \epsilon^i\zeta}}
{ \left(\sum_{i=1}^{i=N}\left(1.28\delta_s^i\right)^2\right)^{1/2}}\hspace{3cm}(\sum\theta_s^i \leq 0)
\label{e:thalflim}
\end{eqnarray}
Here $N_{\rm iso}$ is the number of candidate nuclei in the tin for the given decay, $t_{\rm live}$ is the duration of data collection in years and $\epsilon$ is the efficiency for observing the given gamma signal determined from simulations. All of these parameters are calculated separately for each run and summed over all $N$ runs. $\zeta$ is the branching fraction for the gamma-line searched for. \par
%
Possible systematic effects which could effect the half-lives are discussed in the following. The energies of all the peaks studied are known precisely due to accurate knowledge of the involved gamma energies ($<0.05$\,keV uncertainty). The only search needing special attention is the $0\nu$ \ecec ground state transition as the Q-value for this decay is only known to about 2\,keV due to uncertain atomic mass determinations. This will be discussed in the corresponding section. The total amount of tin is known to a precision of 0.01\,g and thus any associated error is negligible. Furthermore the natural abundances of tin isotopes given in \cite{toi} were used. The tin is pure to 99.95\,\%  and contaminations can only contribute 0.05\,\% which is considered to be small. Dead time corrections are less than 1\,\% throughout due to the low count rates. \par
\subsection{Excited state transitions of \tinhv}
\begin{figure}
\begin{center}
\epsfig{file=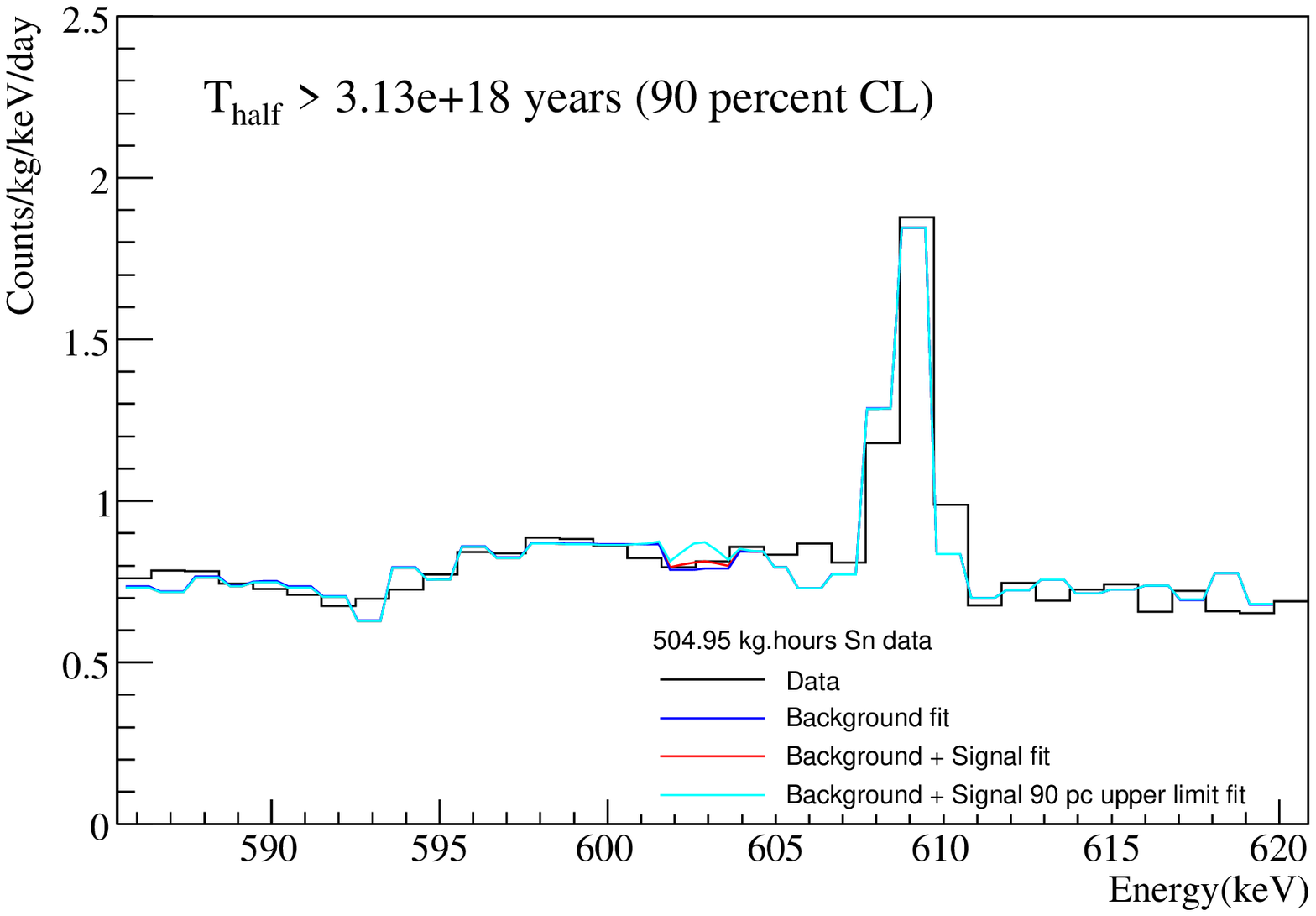,width=12cm,height=8cm}
\end{center}
\caption{\label{pic:lineat603} Summed data and fit results in the region around the 602.7\,keV peak, expected for all decay modes of $^{124}$Sn. The peak at 609.3\,keV is from $^{214}$Bi decay.}
\end{figure}

From the known level scheme of \tehv it can be seen that all higher level decays occur dominantly via the intermediate $2_1^+$-state at 602.7\,keV. Hence a first step is to look for the 602.7\,keV photon of this decay. A magnified region of the energy range of interest from Run 6 is shown in Fig.~\ref{pic:lineat603} along with the fitted gaussian. The energy resolution at this point was determined to be 1.35\,keV (FWHM). With $14 \pm 28$ possible events in the gaussian peak and an efficiency of 1.33~\% a half-life of 
\be
\ton^{0(2)\nu} ( 2^+_1 (602.7\,keV)\ra 0^+ ) > 3.1~(3.7) \times 10^{18}~\mbox{yrs}~(90~(68)~\% CL) 
\ee
has been obtained, an improvement on the limit given in \cite{nor87}. Note that this limit was obtained from Run 6 only, by fitting the measured background spectrum to account for the neutron interaction feature above 596\,keV.\par
The search for decays into higher $0^+$-states involves two gammas. Due to the de-excitation sequence $0^+ \ra 2^+ \ra 0^+$ a strong angular correlation between the two photons is expected. Thus it is very unlikely that both gammas will be simultaneously detected in the Ge-detector and additional searches have been performed for the second gamma energies accompanying the 602.7\,keV. The corresponding half-life limits and efficiencies are compiled in Tab.~\ref{tab:exstate-tinhv}. The calculation of limits takes the branching ratio for the de-excitation chain into account as this can be less than 100\% for the higher level states. For decays to some of the higher level states, the best limit on half-life is derived from one of the lower de-excitation steps in the decay chain (indicated in brackets in the table).
\begin{table}[htbp]
\begin{center}
{\small
\begin{tabular}{|c|c|c|c|c|c|}
\hline
Excited state  & Gamma energy(ies) & Efficiency & Events & \ton (90\% CL)\\
& (keV) & (\%) & (kg$^{-1}$hr$^{-1}$) & ($10^{18}$ yrs)\\
\hline
$2^+_1$ (602.7) & {\bf602.7} & 1.327 $\pm$ 0.012 & $14\pm28^\dag$ &  3.1$^\dag$\\
$0^+_1$ (1156.5) & {\bf553.8}, 602.7 & 1.410 $\pm$ 0.012 & $-27\pm28$ & 7.7 \\
$2^+_2$ (1325.5) & {\bf722.9}, 602.7 & 1.145 $\pm$ 0.011 & $0.7\pm22$ & 4.4\\
$0^+_2$ (1656.7) & {\bf1054.0}, 602.7 & 0.839 $\pm$ 0.009& $-16\pm14$ & 7.9\\
$0^+_3$ (2020.0) & {\bf695.0}, 722.9, 602.7& 1.175 $\pm$ 0.011  & $20\pm27^\dag$ & 2.5$^\dag$ (4.4)\\
$2^+_2$ (2039.3) & {\bf713.8}, 722.9, 602.7& 1.151$ \pm$ 0.011& $-23\pm 23$ & 3.1 (4.4)\\
$2^+_3$ (2091.6) & {\bf1488.9}, 602.7& 0.637 $\pm$ 0.009& $25\pm 10$ & 2.7 (3.1$^\dag$)\\
\hline
\end{tabular}
}
\caption{\label{tab:exstate-tinhv}Half-life limits (90 \% CL) for all possible excited $0^+,2^+$-state transitions (0$\nu$ and 2$\nu$) of $^{124}$Sn.
Given are the excited states, all possible gamma lines with the dominant mode searched for in bold, efficiencies for $\gamma$-detection in the configuration for runs 4--6, the total number of fitted events in the peak and deduced half-life limits. Where the best limit is derived from another step in the decay chain, it is given in brackets. Efficiencies are given for the longest run period. Decays marked \dag~were analysed for Run 6 only, by fitting the measured background spectrum.}
\end{center}
\end{table}
\subsection{Excited state transitions of \tinhz}
\begin{figure}
\begin{center}
\epsfig{file=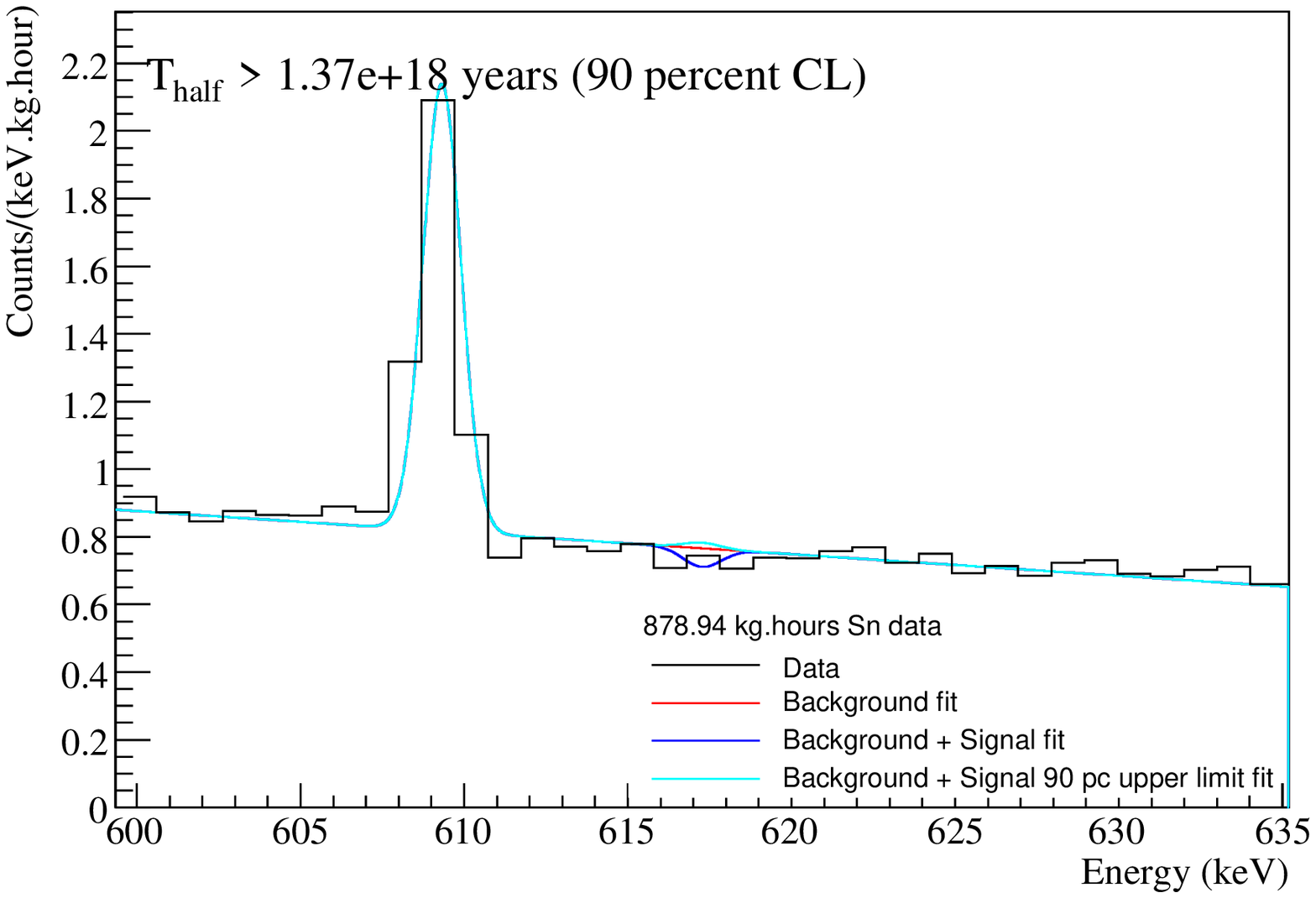,width=12cm,height=8cm}
\end{center}
\caption{\label{pic:peak617}  Summed data and fit results in the region around the 617.3\,keV peak, expected for all decay modes of $^{112}$Sn. The peak at 609.3\,keV is from $^{214}$Bi decay. }
\end{figure}

\begin{figure}
\begin{center}
\epsfig{file=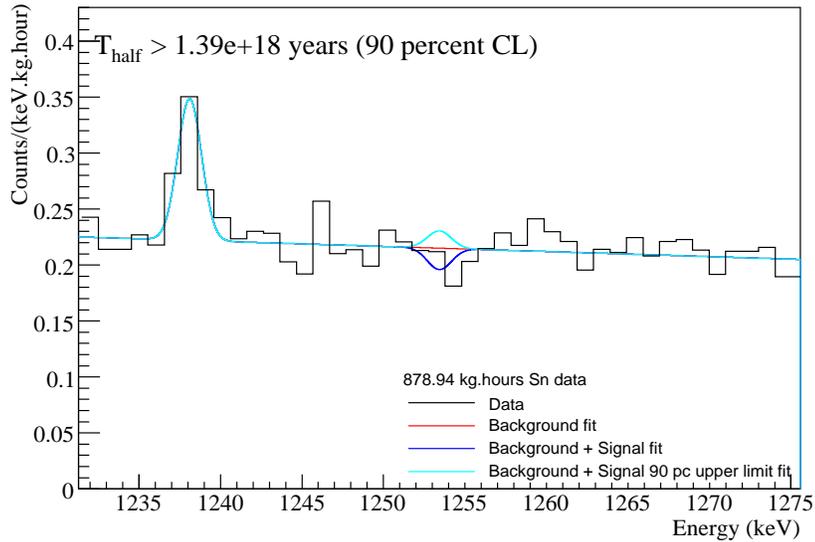,width=12cm,height=8cm}
\end{center}
\caption{\label{pic:peak1253}  Summed data and fit results in the region around the 1253.5\,keV peak expected for the decay of $^{112}$Sn via the 1879.9\,keV $0^+$ excited state that could be resonantly enhanced. The peak at 1238.1\,keV is from the decay of $^{214}$Bi.}
\end{figure}

A rich variety of decays can be studied for \tinhz in the form of \ecec and \bec modes. First of all it should be mentioned that all decays into higher excited states de-excite via a $2_1^+$ state at 617.6\,keV. Thus, a search for this line places a half-life limit on all excited state transitions, taking into account the branching ratios of the de-excitation. No peak is observed as can be seen in Fig.~\ref{pic:peak617} and a limit of 
\be
\label{eq:firstex}
\ton^{0(2)\nu EC/EC + \beta^+/EC} ( 2^+_1 (617.6 {\rm~keV})\ra 0^+)> 1.4\times 10^{18}~\mbox{yrs}~(90\% CL)
\ee
has been derived. A similar search was performed for all potential gamma-lines involved and their half-life limits are compiled in Tab.~\ref{tab:exstate-tinhz}. Of special interest is a $0^+$-state at 1870.9\,keV. This state could be degenerate with the ground state of \tinhz and thus has the possibility of a resonantly enhanced neutrinoless \ecec rate. The de-excitation of this state to ground is dominated by the emission of a 1253.4\,keV and 617.6\,keV gamma. At both gamma energies a search was performed. No obvious peak was observed at either position, the 1253\,keV region is shown in Fig.~\ref{pic:peak1253}, and the corresponding half-life limit for this decay mode is 
\be
\ton^{0(2)\nu EC/EC} ( 0^+(1870.9 {\rm~keV}) \ra 0^+) > 1.6\times 10^{18} \mbox{yrs}~ (90\% CL).
\ee
The \bec modes can only populate excited states up to Q-1.022\,MeV, which in the case of \tinhz means states below 900\,keV excitation energy. The only level available is the $2^+_1$-state at 617.6\,keV. Thus, for the 0(2)$\nu$ \bec-decay modes into the first excited $2^+_1$ state the same half-life limit applies as the one from the 617.6\,keV line search for \mbox{$EC/EC$}, given in Eqn.~\ref{eq:firstex}.  
\begin{table}[htbp]
\begin{center}
{\small
\begin{tabular}{|c|c|c|c|c|c|}
\hline
Excited state  & Gamma energy & Efficiencies & Events   & \ton (90\% CL)\\
& (keV) & (\%) & (kg$^{-1}$hr$^{-1}$) & ($10^{18}$ yrs)\\
\hline
$2^+_1$ (617.6) & {\bf617.6} & 1.296 $\pm$ 0.011 & $-48\pm24$ & 1.4\\
$0^+_1$ (1224.1) &  {\bf606.5}, 617.6  & 1.320 $\pm$ 0.011 & $87\pm28^\dag$ &  0.21$^\dag$ (1.4)\\
$2^+_2$ (1312.3) & {\bf694.7}, 617.6 & 1.175 $\pm$ 0.011 & $14\pm26^\dag$ & 0.35$^\dag$ (1.4)\\
$0^+_2$ (1433.2) & {\bf815.8}, 617.6 & 1.014 $\pm$ 0.010 & $0.6\pm19$ & 0.62 (1.4)\\
$2^+_3$ (1468.7) & {\bf851.1}, 617.6 & 0.988 $\pm$ 0.010 & $5\pm18$ & 0.58 (1.4)\\
$0^+_3$ (1870.9) & {\bf1253.4}, 617.6 & 0.728 $\pm$ 0.009 & $-17\pm12$ & 1.4\\
\hline
\end{tabular}
}
\caption{\label{tab:exstate-tinhz}Half-life limits (90 \% CL) for excited state transitions (0$\nu$ and 2$\nu$) of $^{112}$Sn. Given are the excited states, all possible gamma lines with the dominant mode searched for in bold, efficiencies for $\gamma$-detection in the configuration for runs 4--6, the total number of fitted events in the peak and deduced half-life limits. Where the best limit is derived from another step in the decay chain, it is given in brackets. Efficiencies are given for the longest run period. A \dag~indicates that the limit was obtained by fitting the background spectrum to the data for Run 6 only.}
\end{center}
\end{table}

\subsection{Ground state transitions of \tinhz}
\begin{figure}
\begin{center}
\epsfig{file=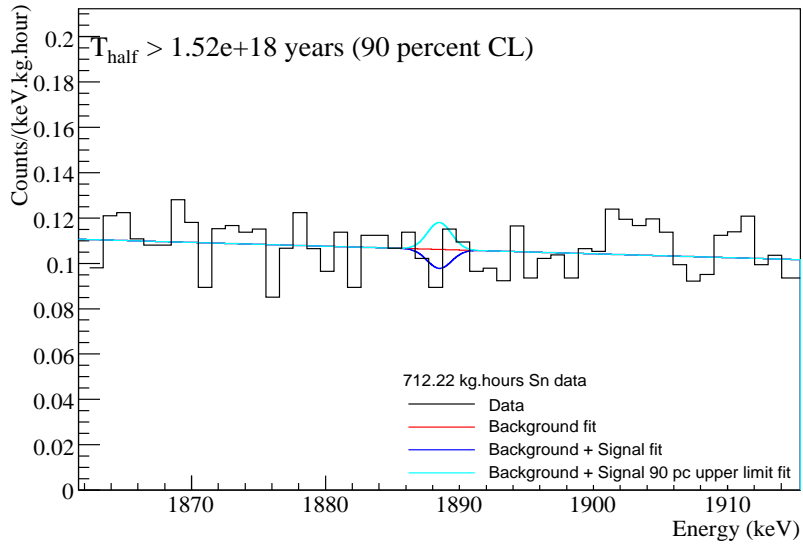,width=12cm,height=8cm}
\end{center}
\caption{\label{pic:peak1888}   Summed data and fit results in the region around the 1888.5\,keV peak which could be expected from a double beta decay of $^{112}$Sn to the ground state. Note only the last 3 runs contribute to this summed spectra due to the upper energy threshold requirements.  }
\end{figure}

Two possible decay modes are considered for the ground state transition. The first one is for the $0\nu$ \ecec ground state transition based on the emission of an internal bremsstrahlung photon \cite{doi93}. Due to energy and momentum conservation this requires one K- and one L-shell capture. Additionally the bremsstrahlung photon has to be monoenergetic, with an energy of the Q-value reduced by the K- and L-shell binding energy of the daughter. In the case of \tinhz this implies a gamma of 1888.5\,keV. The region around this energy is shown in Fig.~\ref{pic:peak1888}. The calculated detection efficiency for such a photon is 0.491 $\pm$ 0.007. The analysis follows the method described for the excited states and due to non-observation of a peak results in a half-life of T$_{1/2}>1.5\times 10^{18}$\,\mbox{yrs} at the 90\% confidence level.
As the precise peak position is only known to about $\pm$ 2\,keV due to uncertainties in the atomic masses, the peak position was systematically varied by 0.5\,keV between 1886 and 1890\,keV and the fit was repeated for the corresponding position. In this way, the worst half-life limit was found to be 
\be
\ton^{0\nu EC/EC} ( 0^+ \ra 0^+_{g.s}) > 9.9\times 10^{17} \mbox{yrs}~(90\% CL),
\ee
which thus can be used as a conservative lower limit for the decay. 

The second search on the 0$\nu$ \bec ground state transition is based on the creation of 511\,keV annihilation gammas. However, there is a difference in the search strategy because an excess of an observable background line at 511\,keV has to be explored. A possible difference in the intensity of the 511\,keV might come from internal impurities of the tin as well as due to the fact that 511\,keV gammas produced externally can be shielded by the tin. As Fig.~\ref{f:background} shows, the continuum level is actually lower in Run 6 with the tin than without. The amplitude of the 511\,keV peak was determined for the attenuated background spectrum, (fitted via Eqn.~\ref{e:atten} as described earlier) and for the Run 6 spectrum with tin. In these fits, the width of the gaussian used to fit the peak was also allowed to vary since 511\,keV emissions can be subject to Doppler broadening ($\sigma \approx 1$\,keV wider) and hence the resolution of this line cannot be accurately calibrated.\par
\begin{figure}
\begin{center}
\epsfig{file=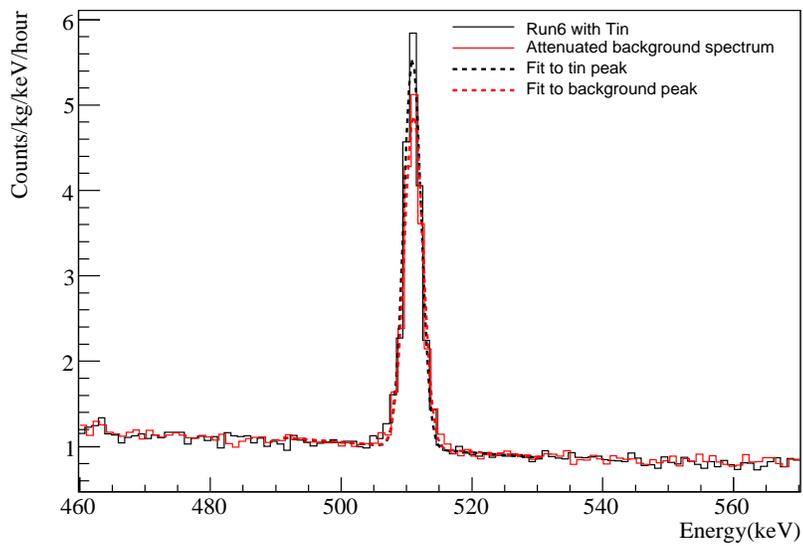,width=12cm,height=8cm}
\end{center}
\caption{\label{pic:peak511} Run 6 data and the attenuated background spectrum around the 511\,keV peak. The gaussian functions fitted to each peak show an excess of events in the data obtained with the tin present.}
\end{figure}

The region fitted around the 511\,keV peak is shown in Fig.~\ref{pic:peak511} for both the tin and background spectra.  The total number of events in the background peak was found to be $1579\pm34$ and for the tin spectrum the total number of events in the peak was $1849\pm37$. The difference between these values, $270\pm50$ events is inconsistent with zero indicating a source of 511\,keV gammas is present in the tin. As a cross-check the same procedure was repeated for other background lines (351.9, 583.2, and 609.3\,keV) and no excess was found in any of these other peaks, so the intensity increase of the 511\,keV line cannot be attributed to uranium or thorium contamination in the tin.\par
Assuming the observed events are all due to the double beta decay of $^{112}$Sn, a half-life of \\
\be
\ton^{0(2)\nu \beta^+/EC} ( 0^+ \ra 0^+_{g.s}) = 5.09^{+1.17}_{-0.79}\times 10^{17} \mbox{yrs}
\ee
can be derived, given the efficiency for observing 511\,keV gammas =1.175\%.\par
However, the introduction of an additional positron source in the tin cannot be ruled out. Further work is required to study possible $\beta^+$ emitters that could be produced from cosmic-ray interactions or neutron activation in tin. Under the assumption that the observed excess of events are all background, a conservative limit of 
\be
\ton^{0(2)\nu \beta^+/EC} ( 0^+ \ra 0^+_{g.s}) > 4.1\times 10^{17} \mbox{yrs}~(90\% CL)
\ee
has been deduced.
\section{Summary and conclusions}
Double beta decay, in its various forms, is a powerful tool for investigating neutrino properties. Many isotopes are being considered for large scale experiments, but so far tin has received little attention. In addition to \tinhv which is suitable for $0\nu\beta^-\beta^-$ searches, tin contains \tinhz available for decay via the \bec mode, which has never been explored experimentally. This paper details the first comprehensive search for excited state transitions in tin isotopes and leads to half life limits in the region of $10^{17}$--$10^{19}$\,yrs. Ground state transitions for \tinhz were explored for neutrinoless radiative \ecec producing a mono-energetic photon. The \bec ground state transition could be restricted observing the 511\,keV line. A small excess of the line is seen, which results in a rather short half-life if all the excess events are attributed to this signal. However, due to unknown activities in the tin produced by cosmic ray spallation and neutron reactions it is more reliable to quote a limit. Improvements could be obtained with more sophisticated shielding and a better detection efficiency.

\section{Acknowledgement}
We thank Nick Jelley (University of Oxford) for providing clean copper and lead parts. J.R. Wilson is supported by a PPARC fellowship.

\end{document}